\documentclass[aps,nofootinbib]{revtex4-2}

\usepackage{geometry}                
\usepackage{amsmath,amssymb,amsfonts,dcolumn,color,graphicx}
\usepackage{latexsym,placeins,epsfig,tikz,amsbsy,bm,enumitem}
\usetikzlibrary{arrows,shapes}
\usepackage{subfigure,rotating,bm,mathrsfs}
\usepackage[title]{appendix}
\usepackage[pagebackref=false, colorlinks=true]{hyperref}
\hypersetup{linkcolor=red,         
                  citecolor=blue,        
                  filecolor=magenta,      
                  urlcolor=blue}          

\newcommand{\calH}{{\cal H}}

\allowdisplaybreaks

\begin{document}

\rightline{APCTP-Pre2022-013}

\vspace{1em}

\title{\Large Second-order energy-momentum tensor of a scalar field}

\author{Inyong Cho}
\email{iycho@seoultech.ac.kr}
\affiliation{School of Natural Sciences, College of Liberal Arts,
Seoul National University of Science and Technology, Seoul 01811, Korea}
\author{Jinn-Ouk Gong}
\email{jgong@ewha.ac.kr}
\affiliation{Department of Science Education,  Ewha Womans University, Seoul 03760, Korea \\
Asia Pacific Center for Theoretical Physics,  Pohang 37673, Korea}
\author{Seung Hun Oh}
\email{sho@tukorea.ac.kr}
\affiliation{Department of Consilience, Tech University of Korea, Siheung 15073, Korea}

\begin{abstract}

We investigate the second-order effective energy-momentum tensor (2EMT) constructed by
the quadratic terms of the linear scalar cosmological perturbations  
while the universe is dominated by a scalar field.
We show that 2EMT is gauge dependent.
We then study 2EMT in three (longitudinal, spatially flat, and comoving) 
gauge conditions in the slow-roll stage of inflation.
We find that 2EMT exhibits an effective fluid of $w=-1/3$ on super-horizon scales
in all of those gauge conditions.

\end{abstract}

\maketitle


\section{Introduction}

Inflationary universe is a successful scenario so far,
according to the observation of the cosmic  microwave background (CMB).
Cosmological perturbations produced during the inflationary period explain the CMB data successfully.
As the precision of the observational technique improves,
the higher-order perturbations beyond the linear order have attracted more attentions. 
Thus, it is increasingly more important to understand higher-order nature of cosmological perturbations 
to test the paradigm of inflation and to discriminate different models of inflation.
The second-order cosmological perturbations during inflation
were investigated in Refs.~\cite{Mukhanov:1996ak,Abramo:1997hu}, 
in particular, 
the second-order effective energy-momentum tensor (2EMT) constructed by the quadratic terms 
of the linear perturbations was considered.
This has been studied as a way of backreaction of perturbations on the background.
It was also discussed the gauge invariance of 2EMT.

Recently 2EMT of cosmological perturbations produced by perfect fluid was investigated 
in the Friedmann universe \cite{Cho:2020zbh}.
The gauge dependence of 2EMT was presented in general.
The 2EMT in three gauge conditions (longitudinal, spatially-flat, and comoving gauges)
was investigated in the matter- and radiation-dominated epochs.
No convergence of 2EMT was observed,
i.e., each gauge condition gives a different 2EMT in the given epoch.

The story of the gauge invariance of the Einstein's equation can be discussed as below.
One can expand the Einstein's equation by order as (let $8\pi G=1$ in this section),
\begin{align}
\begin{split}
G_{\mu\nu} &= T_{\mu\nu}  
\\
\quad\Rightarrow\quad
G^{(0)}_{\mu\nu} + G^{(1)}_{\mu\nu} + G^{(2)}_{\mu\nu} + \cdots
&= T^{(0)}_{\mu\nu} + T^{(1)}_{\mu\nu} + T^{(2)}_{\mu\nu} + \cdots .
\end{split}
\end{align}
The equality is satisfied order by order, $G^{(n)}_{\mu\nu} = T^{(n)}_{\mu\nu}$.
Let us consider the first-order equation,  $G^{(1)}_{\mu\nu} = T^{(1)}_{\mu\nu}$.
The Einstein tensor $G^{(1)}_{\mu\nu}$ and the energy-momentum tensor $T^{(1)}_{\mu\nu}$
are not gauge invariant by themselves.
By adding terms on both sides of  the first-order equation,
one can construct the equation in a gauge invariant form,
\begin{align}
\overline{G}^{(1)}_{\mu\nu} = \overline{T}^{(1)}_{\mu\nu},
\end{align} 
where the gauge invariant quantities are given by
\begin{align}
\overline{G}^{(1)}_{\mu\nu} = G^{(1)}_{\mu\nu} + \delta G^{(1)}_{\mu\nu},
\qquad
\overline{T}^{(1)}_{\mu\nu} = T^{(1)}_{\mu\nu} + \delta T^{(1)}_{\mu\nu}.
\end{align} 
Here, $\delta G^{(1)}_{\mu\nu}$ and $\delta T^{(1)}_{\mu\nu}$ are the terms added
to make gauge invariant quantities.

Similarly, the second-order equation can be put into the gauge invariant form,
$\overline{G}^{(2)}_{\mu\nu} = \overline{T}^{(2)}_{\mu\nu}$.
However, each quantity is composed in a more complicated way,
\begin{align}
\overline{G}^{(2)}_{\mu\nu} = G^{(2)}_{\mu\nu} + \delta G^{(2)}_{\mu\nu}
& = G^{(1)}_{\mu\nu}[g^{(2)}] + \underline{G^{(2)}_{\mu\nu}[g^{(1)}]}
+\delta G^{(1)}_{\mu\nu}[g^{(2)}] + \delta G^{(2)}_{\mu\nu}[g^{(1)}]
\label{G2gi}
\\
\overline{T}^{(2)}_{\mu\nu} = T^{(2)}_{\mu\nu} + \delta T^{(2)}_{\mu\nu}
& = T^{(1)}_{\mu\nu}[g^{(2)},\psi^{(2)}] + \underline{T^{(2)}_{\mu\nu}[g^{(1)},\psi^{(1)}]}
+\delta T^{(1)}_{\mu\nu}[g^{(2)},\psi^{(2)}] + \delta T^{(2)}_{\mu\nu}[g^{(1)},\psi^{(1)}] ,
\label{T2gi}
\end{align}
where $g^{(n)}$ and $\psi^{(n)}$ represent the metric and the matter perturbations of $n$-th order.
The second-order quantity consists of two parts;
one is linear order of the second-order perturbations (e.g., $G^{(1)}_{\mu\nu}[g^{(2)}]$),
and the other is quadratic-order of the linear perturbations (e.g., $G^{(2)}_{\mu\nu}[g^{(1)}]$).

2EMT is constructed by the underlined terms in Eqs.~\eqref{G2gi} and \eqref{T2gi},
\begin{align}
T^{(2,{\rm  eff})}_{\mu\nu} \equiv 
T^{(2)}_{\mu\nu}[g^{(1)},\psi^{(1)}] - G^{(2)}_{\mu\nu}[g^{(1)}].
\end{align}
Considering $\overline{G}^{(2)}_{\mu\nu}$  and $\overline{T}^{(2)}_{\mu\nu}$
are gauge invariant individually,
it is hard to expect
$T^{(2,{\rm  eff})}_{\mu\nu}$ which is only a collection of parts of them
to be gauge invariant.

In Ref.~\cite{Cho:2020zbh}, we derived 2EMT of barotropic fluid,
and showed that it is indeed gauge dependent.
In this work, we study 2EMT of a minimally coupled canonical scalar field.
We will derive 2EMT and show its gauge dependence.
We shall try to impose three gauge conditions (longitudinal, spatially-flat, and comoving gauges),
and investigate the convergence of 2EMT in the slow-roll background during inflation. 
Very interestingly, we find that 2EMT in all gauge choices converges to an effective fluid
in the long-wavelength limit. This convergence is absent in the barotropic fluid 
in the Friemann universe studied in our previous work~\cite{Cho:2020zbh}.

\section{Einstein's equation and 2EMT of scalar field}
\label{sec:2emt}

In this section, we derive 2EMT starting from the linear perturbations.
We consider the general metric 
\begin{align}
\label{metric1}
ds^{2} 
= a^{2}(\eta) \Big[ -(1+2A)d\eta^{2} - 2 B_{i} d\eta dx^{i}
+ (\delta_{ij} + 2 C_{ij}) dx^{i} dx^{j} \Big] 
  ,
\end{align}
where the metric perturbations $A$, $B_i$, and $C_{ij}$ 
are to be expanded in all orders.
The Einstein tensor $G_{\mu\nu}$ up to second order was presented in our previous work~\cite{Cho:2020zbh}, 
which we will not duplicate here.
In this section, we  present the energy-momentum tensor of a minimally coupled canonical scalar field $\phi$.

\subsection{Energy-momentum tensor}

The energy-momentum tensor of a minimally coupled scalar field $\phi$ is given by the perturbed matter action $S_\phi$ with respect to the metric $g^{\mu\nu}$,
\begin{equation}
T_{\mu\nu} 
= 
-\frac{2}{\sqrt{-g}} \frac{\delta{S}_\phi}{\delta{g}^{\mu\nu}}
=
\phi_{,\mu}\phi_{,\nu} - g_{\mu\nu} \bigg[ \frac{1}{2} g^{\rho\sigma} \phi_{,\rho} \phi_{,\sigma} - V(\phi) \bigg] 
  ,
\end{equation}
where $V(\phi)$ is the potential of $\phi$. We split $\phi$ into the unperturbed background $\phi_0$ 
and the perturbation $\delta\phi$,
\begin{equation}
\phi = \phi_0 + \delta\phi   .
\end{equation}                    
Then each component of the energy-momentum tensor is readily found as, up to second order in perturbations,
\begin{align}
\label{eq:T00}
T_{00}
& = 
\frac{1}{2}({\phi_{0}}^{\prime})^{2} + a^2 V(\phi_{0})
+ \phi_{0}^{\prime} \delta \phi^{\prime}
+ 2 A V(\phi_{0})
+ a^2V_\phi(\phi_{0}) \delta \phi
\nonumber \\
& \quad
+ \frac{1}{2} (\delta \phi^{\prime})^{2}   
+ \frac{1}{2} (\nabla \delta \phi)^{2} 
+ \frac{1}{2}a^2 V_{\phi\phi}(\phi_{0}) \delta \phi^{2} 
+ 2a^2AV_\phi(\phi_{0})\delta \phi
+ \frac{1}{2} B_{k}B_{k}(\phi_{0}^{\prime})^{2}
- \phi_{0}^{\prime} B_{k} \delta\phi_{,k} 
  ,      
\\
\label{eq:T0i}
T_{0i}
& = 
\phi_{0}^{\prime} \delta\phi_{,i}
- \frac{1}{2} B_{i} (\phi_{0}^{\prime})^{2}
+ a^2 B_{i} V(\phi_{0}) 
\nonumber\\
& \quad
+ \delta \phi^{\prime} \delta \phi_{,i}
- B_{i} \phi_{0}^{\prime} \delta \phi^{\prime} 
+ AB_{i}(\phi_{0}^{\prime})^{2}
+ a^2 B_{i} V_\phi(\phi_{0}) \delta \phi
  ,
\\
\label{eq:Tij}
T_{ij}
& = 
\delta_{ij} \bigg[ \frac{1}{2} (\phi_{0}^{\prime})^{2} - a^{2}V(\phi_{0}) \bigg]
+ \delta_{ij}  \Big[ \phi_{0}^{\prime} - A(\phi_{0}^{\prime})^{2} - a^{2}V^{\prime}(\phi_{0})\delta \phi \Big]
+ C_{ij} \Big[ (\phi_{0}^{\prime})^{2} - 2a^{2}V(\phi_{0}) \Big]
\nonumber \\
& \quad
+ \delta \phi_{,i}\delta \phi_{,j} 
+ 2C_{ij} \Big[ \phi_{0}^{\prime} \delta \phi^{\prime} 
- A (\phi_{0}^{\prime})^{2}
- a^{2}V_\phi (\phi_{0}) \delta \phi  \Big]
\nonumber \\
& \quad
+ \delta_{ij} \bigg[ \frac{1}{2} (\delta \phi^{\prime})^{2} - \frac{1}{2} (\nabla \delta \phi)^{2}
+ \bigg(2A^{2} - \frac{1}{2}B_{k}B_{k} \bigg) (\phi_{0}^{\prime})^{2} 
- 2 A \phi_{0}^{\prime} \delta \phi^{\prime} + B_{k} \phi_{0}^{\prime} \delta \phi_{,k}
-\frac{1}{2}a^{2} V_{\phi\phi} (\phi_{0}) \delta \phi^{2} \bigg] 
  ,
\end{align}
where a prime denotes a derivative with respect to the conformal time $\eta$, 
$V_\phi = \partial{V}/\partial\phi|_0$ and $V_{\phi\phi} = \partial^2V/\partial\phi^2|_0$.
From now on, the potential is always evaluated with respect to the background value, 
i.e., $V = V(\phi_0)$ and so on.

\subsection{Einstein's equation}

We can construct order by order the Einstein's equation 
by equating each component of the Einstein tensor 
and the corresponding component of the energy-momentum tensor \eqref{eq:T00}-\eqref{eq:Tij}.

\subsubsection{Background equations}

The background equations which contain no perturbations but only background variables, 
are given by the $00$- and $ij$-components,
\begin{align}
\label{eq:BGeq1}
\calH^2 & = \frac{8\pi G}{3} \bigg[ \frac{1}{2} (\phi_0')^2 + a^2V \bigg],
\\
\label{eq:BGeq2}
2\calH' + \calH^2 & = 8\pi G \bigg[ -\frac{1}{2} (\phi_0')^2 + a^2V \bigg],
\end{align}
where ${\cal H} \equiv (da/d\eta)/a$, 
and for later use, $H \equiv (da/dt)/a$ with $dt =ad\eta$.
The equation for $\phi_0$ can be found from the energy-momentum tensor conservation,
\begin{equation}
\label{eq:BGeq3}
\phi_0'' + 2\calH\phi_0' + a^2V_\phi = 0   .
\end{equation}

\subsubsection{Linearized equations}

Using the background equations \eqref{eq:BGeq1}-\eqref{eq:BGeq3}, 
we find the most convenient form of the first-order equations,
\begin{align}
\label{eq:lineareq1}
&
2(\mathcal{H}^{\prime} + 2\mathcal{H}^{2}) A  - 2\mathcal{H} B_{k,k} 
- 2 \mathcal{H} C^{\prime}_{kk} + \Delta C_{kk} -C_{kl,kl} 
= 
 8\pi G \Big[ (\phi_{0}^{\prime} \delta \phi)^{\prime} 
+ 2\mathcal{H} \phi_{0}^{\prime} \delta \phi - 2\phi_{0}^{\prime} 
\delta \phi^{\prime} \Big]  
  ,  
\\
\label{eq:lineareq2}
&
2 \mathcal{H} A_{,i} + B_{[i,k]k} + C^{\prime}_{ik,k} - C^{\prime}_{kk,i}
= 
8\pi G \phi_{0}^{\prime} \delta \phi_{,i} 
  ,
\\
\label{eq:lineareq3}
&
\delta_{ij} \Big[ 2\mathcal{H}A^{\prime} 
+ 2 \big( \mathcal{H}^{\prime} + 2 \mathcal{H}^{2} \big) A + \Delta A 
- B_{k,k}^{\prime} - 2 \mathcal{H}B_{k,k} 
- C_{kk}^{\prime \prime} - 2\mathcal{H} C_{kk}^{\prime} + \Delta C_{kk}  - C_{kl,kl} \Big] 
\nonumber  \\
&
- A_{,ij} + B^{\prime}_{(i,j)} + 2\mathcal{H}B_{(i,j)} 
+ C_{ij}^{\prime\prime} + 2\mathcal{H} C_{ij}^{\prime} - \Delta C_{ij} + 2C_{k(i,j)k} - C_{kk,ij} 
\nonumber\\
&
= 8\pi G \delta_{ij} \Big[ (\phi_{0}^{\prime} \delta \phi)^{\prime} 
+ 2\mathcal{H} \phi_{0}^{\prime} \delta \phi  \Big] 
  .   
\end{align}
The equation of motion for the scalar-field perturbation $\delta\phi$ is given by
\begin{equation}
\label{Eq:lineareq4}
\delta \phi^{\prime \prime} + 2\mathcal{H}\delta \phi^{\prime}
-\Delta \delta \phi + a^{2} V_{\phi \phi}  \delta \phi 
- (A^{\prime} - B_{k,k} -C_{kk}^{\prime}) \phi_{0}^{\prime} 
+ 2a^{2} V_{\phi} A = 0 
  .
\end{equation}

\subsubsection{Gauge-invariant variables}

Now, we consider only scalar components of the metric perturbations 
such that they are written in terms of four functions,
\begin{equation}
A = \alpha   ,
\quad
B_i = \beta_{,i}   ,
\quad
C_{ij} = -\psi\delta_{ij} + E_{,ij}   .
\end{equation}
These functions transform under the infinitisimal coordinate transformation $x^\mu \to x^\mu + \xi^\mu$ as,
\begin{align}
\alpha & \to \alpha - {\xi^0}' - \calH\xi^0   ,
\\
\beta & \to \beta - \xi^0 + \xi'   ,
\\
\psi & \to \psi + \calH\xi^0   ,
\\
E & \to E - \xi   .
\end{align}
Then, it can be readily shown that the following variables are gauge-invariant up to first order,
\begin{align}
\overline{\delta \phi} & = \delta \phi - \phi_{0}^{\prime}  Q   ,
\\
\Phi & = \alpha - Q^{\prime} - \mathcal{H} Q   ,
\\
\Psi & = \psi + \mathcal{H} Q   ,
\end{align} 
where $Q$ represents a time translation defined by
\begin{equation}
Q = \beta + E'.
\end{equation}
Now, Eqs.~\eqref{eq:lineareq1}-\eqref{eq:lineareq3} enable us to express 
$\overline{\delta\phi}$ and $\overline{\delta\phi}'$ 
in terms of the gauge-invariant variable $\Psi = \Phi$,
\begin{align}
\overline{\delta \phi} & = \frac{\Psi^{\prime} + \mathcal{H} \Psi}{4\pi G \phi_{0}^{\prime}}
  , 
\\
\overline{\delta \phi}^{\prime} & = \frac{\Delta \Psi - K \Psi^{\prime} - L \Psi }{4\pi G \phi_{0}^{\prime}}
  ,
\end{align}
where $K$ and $L$ depend only on the background variables,
\begin{align}
K & = 3\mathcal{H} + a^{2} \frac{V_{\phi} }{ \phi_{0}^{\prime} }
  ,
\\
L & = \mathcal{H}^{\prime} + 2\mathcal{H}^{2} 
+a^{2} \mathcal{H} \frac{V_{\phi} }{ \phi_{0}^{\prime}}
  .
\end{align}
Then, the linear equations \eqref{eq:lineareq1}-\eqref{eq:lineareq3} 
can be written solely in terms of $\Psi$ as
\begin{align}
\label{eq:Psi-eq}
\Psi^{\prime \prime} - \Delta \Psi + 2 K \Psi^{\prime} + 2L \Psi = 0   .
\end{align}

\subsection{2EMT}

Since we have obtained the Einstein and the energy-momentum tensors up to second order, 
and the Einstein's equation up to linear order, 
we now can construct the 2EMT in a similar manner as we did in our previous work~\cite{Cho:2020zbh}. 
As we discussed in Introduction,
the second-order Einstein's equation is rearranged as
\begin{equation}
G_{\mu\nu}^{(1)}[g^{(2)}]
=
8\pi GT_{\mu\nu}^{(1)}[g^{(2)},\delta\phi^{(2)}]
+ 8\pi G T^{(2,{\rm  eff})}_{\mu\nu}
\quad \text{with} \quad
T^{(2,{\rm  eff})}_{\mu\nu} \equiv 
T_{\mu\nu}^{(2)}[g^{(1)},\delta\phi^{(1)}] - \frac{G_{\mu\nu}^{(2)}[g^{(1)}]}{8\pi G} 
,
\end{equation}
where the two quadratic terms make the 2EMT that describes the {\it backreaction} of the linear perturbations.
Using the results in the previous section, 
we can write $T_{\mu\nu}^{(2,\text{eff})}$ in terms of the gauge-invariant variable $\Psi$ 
and the gauge-dependent variables $Q$ and $E$. 
After taking the spatial average denoted by braket notations, 
$\tau_{\mu\nu} \equiv \left\langle T_{\mu\nu}^{(2,\text{eff})} \right\rangle$ ,
we obtain
\begin{align}
\label{eq:2emt1}
8\pi G  \tau_{00} 
& = \frac{1}{\mathcal{H}^{\prime} -\mathcal{H}^{2} } \bigg\{ 
-\big\langle (\nabla \Psi^{\prime})^{2} \big\rangle 
-\big\langle (\Delta \Psi)^{2} \big\rangle
-2(K +\mathcal{H}) \big\langle \nabla \Psi^{\prime} \cdot 
\nabla \Psi \big\rangle 
- \Big[ 3(\mathcal{H}^{\prime} - \mathcal{H}^{2}) + K^{2} 
+ a^{2} V_{\phi \phi}  \Big] 
\big\langle (\Psi^{\prime})^{2} \big\rangle
\nonumber \\
& \hspace{6em}
+\big( 9\mathcal{H}^{\prime} - 10\mathcal{H}^{2}  
- 2L \big) \big\langle (\nabla \Psi)^{2} \big\rangle    
+ 2 \bigg[ 6\mathcal{H}(\mathcal{H}^{\prime} -\mathcal{H}^{2} ) 
- KL + 2(\mathcal{H}^{\prime} - \mathcal{H}^{2})  a^{2} \frac{V_{\phi}}{\phi_{0}^{\prime}} 
- a^{2}\mathcal{H} V_{\phi\phi}  \bigg] 
 \big\langle \Psi \Psi^{\prime} \big\rangle    
 \nonumber \\
&\hspace{6em}
- \bigg[ L^{2} - 4 \mathcal{H}  (\mathcal{H}^{\prime} - \mathcal{H}^{2})
  a^{2} \frac{V_{\phi}}{\phi_{0}^{\prime}} 
+ a^{2}\mathcal{H}^{2} V_{\phi\phi}  \bigg]
\big\langle \Psi^{2} \big\rangle \bigg\}   
\nonumber \\
& 
\quad
+ 2 \left\langle \bigg\{
a^{2}  \frac{V_{\phi}}{\phi_{0}^{\prime}} Q^{\prime}
+ \bigg[ 3\mathcal{H}^{\prime} + (4 \mathcal{H} + K)  a^{2} \frac{V_{\phi}}{\phi_{0}^{\prime}} 
+ a^{2} V_{\phi \phi}  \bigg] Q  
- \Delta Q + \Delta E^{\prime} - 2 \mathcal{H} \Delta E
\bigg\} \Psi^{\prime} \right\rangle 
\nonumber \\
&
\quad
+ 2 \bigg\langle \bigg\{
- \bigg( L +  6\mathcal{H}^{2} - 2 a^{2} \mathcal{H} \frac{V_{\phi}}{\phi_{0}^{\prime}}  \bigg) Q^{\prime}
+ \bigg[ 2 \mathcal{H} (L-3 \mathcal{H}^{\prime}) 
+ (L - 2 \mathcal{H}^{\prime} + 4 \mathcal{H}^{2} )   a^{2} \frac{V_{\phi}}{\phi_{0}^{\prime}} 
+  \mathcal{H} a^{2} V_{\phi \phi}   \bigg] Q
\nonumber \\
&
\hspace{4em}
- \Delta Q^{\prime}  - (K -  \mathcal{H}) \Delta Q
+ 2 \mathcal{H} \Delta E^{\prime} + \Delta^{2} E 
\bigg\} \Psi 
\bigg\rangle 
-  (\mathcal{H}^{\prime} + 2 \mathcal{H}^{2}) \big\langle Q^{\prime 2} \big\rangle 
- 2 \bigg[ \mathcal{H}  (\mathcal{H}^{\prime} - 4 \mathcal{H}^{2}) 
+  (\mathcal{H}^{\prime} 
-\mathcal{H}^{2})  a^{2} \frac{V_{\phi}}{\phi_{0}^{\prime}}   \bigg] 
 \big\langle Q^{\prime}Q \big\rangle 
\nonumber \\
&
\quad
- \bigg\{ (\mathcal{H}^{\prime} - \mathcal{H}^{2}) 
\bigg[ 4 \mathcal{H}^{2} + 8 \mathcal{H}  a^{2}  \frac{V_{\phi}}{\phi_{0}^{\prime}} 
+ a^{4}  \bigg( \frac{V_{\phi}}{\phi_{0}^{\prime}} \bigg)^{2} 
+   a^{2}  V_{\phi \phi}  \bigg] 
+ 3 \mathcal{H}^{\prime} (\mathcal{H}^{\prime} - 4\mathcal{H}^{2}) 
\bigg\} 
\big\langle Q^{2}\big\rangle 
\nonumber \\
&
\quad
- 2 \mathcal{H} \big\langle  \nabla Q^{\prime} \cdot \nabla Q  \big\rangle
+ \big\langle \big[  4 \mathcal{H}^{2} Q + 2 \mathcal{H} \Delta E
- (\mathcal{H}^{\prime} + 2 \mathcal{H}^{2})  E^{\prime}
\big] \Delta E^{\prime}   \big\rangle
+ 4 \mathcal{H} \big\langle (\mathcal{H} Q^{\prime} 
+ \mathcal{H}^{\prime}Q) \Delta E \big\rangle 
  , 
\\
\label{eq:2emt2}
8\pi G\tau_{0i} & = 0  
  ,
\\
\label{eq:2emt3}
8\pi G \tau_{ij} 
& = 
\delta_{ij} \Bigg[
\frac{1}{ \mathcal{H}^{\prime} -\mathcal{H}^{2} } 
\bigg\{ \frac{1}{3}\big\langle (\nabla \Psi^{\prime})^{2} \big\rangle 
-  \big\langle (\Delta \Psi)^{2} \big\rangle
+ 2\bigg(\frac{\mathcal{H}}{3} - K \bigg)
\big\langle \nabla \Psi^{\prime} \cdot \nabla \Psi \big\rangle 
+ \Big( \mathcal{H}^{\prime} - \mathcal{H}^{2} - K^{2} + a^{2} V_{\phi \phi} \Big) 
\big\langle (\Psi^{\prime})^{2} \big\rangle
\nonumber \\
& \hspace{6em}
+\bigg[ \frac{1}{3} \big( 11\mathcal{H}^{\prime} - 10\mathcal{H}^{2} \big) - 2L \bigg] 
\big\langle (\nabla \Psi)^{2} \big\rangle    
+ 2 \Big[  \mathcal{H} a^{2} V_{\phi\phi}  - KL
+ 2 (\mathcal{H}^{\prime} -\mathcal{H}^{2} ) ( K + 2 \mathcal{H} )  \Big] 
 \big\langle \Psi \Psi^{\prime} \big\rangle    
 \nonumber \\
& \hspace{6em}
+ \Big[ 4 ( \mathcal{H}^{\prime} - \mathcal{H}^{2}) 
(\mathcal{H}^{\prime} +2\mathcal{H}^{2} + L) 
- L^{2}  + \mathcal{H}^{2} a^{2} V_{\phi\phi}  \Big]
\big\langle \Psi^{2} \big\rangle 
\bigg\}   
\nonumber \\
& 
\quad
+ 2 \bigg\langle \bigg[  
Q^{\prime \prime} + 3 (3\mathcal{H} - K) Q^{\prime}
 + \bigg( \mathcal{H}^{\prime} + 12 \mathcal{H}^{2}
 + K^{2} - 7 \mathcal{H} K - a^{2} V_{\phi\phi} \bigg) Q
 + \frac{1}{3} \Delta E^{\prime} - \frac{4}{3} (K- \mathcal{H}) \Delta E
\bigg]
\Psi^{\prime} 
\bigg\rangle 
\nonumber \\
& 
\quad
+ 2 \bigg\langle 
\bigg\{ 4 \mathcal{H} Q^{\prime \prime} 
+ ( 6 \mathcal{H}^{\prime} + 10  \mathcal{H}^{2} - 3 L )Q^{\prime}
+ \Delta Q^{\prime} -  ( K - 3 \mathcal{H} ) \Delta Q 
\nonumber \\
& \hspace{3em}
- \Big[  2 \mathcal{H}^{\prime \prime} 
+ a^{2} \mathcal{H} V_{\phi\phi} 
- \mathcal{H} K^{2} + ( 5\mathcal{H}^{\prime} + 2\mathcal{H}^{2}) K
- 11\mathcal{H}^{\prime} \mathcal{H} - 9 \mathcal{H}^{3} \Big] Q
- \frac{1}{3} \Delta E^{\prime \prime} - \frac{2}{3} \mathcal{H} \Delta E^{\prime} 
- \frac{4}{3} L \Delta E + \frac{2}{3} \Delta^{2} E 
\bigg\} \Psi  
\bigg\rangle 
\nonumber \\
& 
\quad
+ \frac{2}{3}\big\langle \nabla Q^{\prime \prime} \cdot \nabla Q  \big\rangle 
+ \frac{2}{3} \big\langle (\nabla Q^{\prime})^{2}  \big\rangle  
+  \frac{4\mathcal{H}}{3} \big\langle \nabla Q^{\prime} \cdot \nabla Q  \big\rangle   
+ 2 \mathcal{H} \big\langle Q^{\prime \prime} Q^{\prime}   \big\rangle 
- 2 \mathcal{H}^{\prime} \big\langle Q^{\prime \prime} Q  \big\rangle 
-  ( \mathcal{H}^{\prime} - 2 \mathcal{H}^{2} )  \big\langle Q^{\prime 2}  \big\rangle  
\nonumber \\
& 
\quad
- 2 \Big[ 2    \mathcal{H}^{\prime \prime}  + 3 \mathcal{H}^{\prime} \mathcal{H}
+  ( \mathcal{H}^{\prime} - \mathcal{H}^{2} )  ( K + \mathcal{H})
\Big]  \big\langle Q^{\prime}Q  \big\rangle 
\nonumber \\
& 
\quad
- \left\{  4\mathcal{H}\mathcal{H}^{\prime \prime} + 3 \mathcal{H}^{\prime 2} 
+ 4 \mathcal{H}^{\prime} \mathcal{H}^{2}  + (\mathcal{H}^{\prime} - \mathcal{H}^{2}) 
\left[ 4 \mathcal{H}^{2} + a^{2} V_{\phi \phi} 
+ a^{4}  \bigg( \frac{V_{\phi}}{\phi_{0}^{\prime}} \bigg)^{2}  \right] 
\right\} \big\langle Q^{2} \big\rangle 
\nonumber \\
& 
\quad
- \frac{2}{3} \big\langle  \big(  2 \mathcal{H} Q 
-3\mathcal{H} E^{\prime} +  \Delta E  \big) \Delta E^{\prime \prime}  \big\rangle 
- \frac{1}{3} \Big\langle  \Big[  
 8 \mathcal{H} Q^{\prime} + 8(\mathcal{H}^{\prime} + \mathcal{H}^{2})Q
-3(\mathcal{H}^{\prime} + 2 \mathcal{H}^{2}) E^{\prime}
+ 2 \Delta E^{\prime} + 4\mathcal{H} \Delta E
\Big] \Delta E^{\prime}  \Big\rangle 
\nonumber \\
& 
\quad
- \frac{4}{3} \Big\langle  \Big\{  
 \mathcal{H} Q^{\prime \prime} + 2(\mathcal{H}^{\prime} +\mathcal{H}^{2})Q^{\prime} 
+ 2  \Big[ 3 \mathcal{H} \mathcal{H}^{\prime} - \mathcal{H}^{3}
- (\mathcal{H}^{\prime} - \mathcal{H}^{2})K \Big] Q  \Big\} \Delta E  \Big\rangle  
\Bigg]
  .
\end{align} 
As we can see, 2EMT does not depend only on the gauge-invariant variable $\Psi$, 
but also on the gauge-dependent variables $Q$ and $E$.
Therefore, we conclude that  the 2EMT for a scalar field is {\it gauge dependent}
like for perfect fluid in Ref.~\cite{Cho:2020zbh}.

\section{2EMT in different gauges}

In the previous section, we have seen that $\tau_{\mu\nu}$ for a scalar field is gauge dependent. 
Still, we may expect that even under different gauge conditions $\tau_{\mu\nu}$ behaves similarly 
in certain wavelength limits so that $\tau_{\mu\nu}$ describes effectively a well-behaved fluid independent of gauge choices. 
In this section, we examine this possibility under three gauge conditions: longitudinal, spatially flat and comoving gauges. 
We consider the long- and short-wavelength limits of $\tau_{\mu\nu}$ in each gauge. 
Here, the long-wavelength limit means the physical wavelengths $\lambda_\text{phys}$ of the perturbation modes 
are much larger than the Hubble scale $H^{-1}$, $\lambda_{\rm phys} \gg H^{-1}$, 
or in terms of the (comoving) momentum,
\begin{equation}
k \ll \calH   .
\end{equation}
In the short-wavelength limit, conversely $k \gg \calH$.

Since we are interested in the Universe dominated by a scalar field, 
i.e., the inflationary epoch representatively, 
we expand $\tau_{\mu\nu}$ in the power of the slow-roll parameters,
\begin{align}
\epsilon & \equiv -\frac{\dot{H}}{H^2} = 4\pi G \frac{\dot\phi_0^2}{H^2} 
  ,
\\
\delta & \equiv -\frac{\ddot\phi_0}{H\dot\phi_0} = \epsilon - \frac{\dot\epsilon}{2H\epsilon} 
  .
\end{align}
With the Fourier-mode expansion 
\begin{equation}
\label{eq:Fourier-mode}
\Psi (\eta,{\bm x}) = \sum_{{\bm k}}  {\Psi}_{{\bm k}}(\eta) e^{i {\bm k} \cdot {\bm x}}   ,
\end{equation}
solving Eq.~\eqref{eq:Psi-eq} in the long- and short-wavelength limits 
gives the following solutions,
\begin{equation}
\label{eq:Psi-sol}
\Psi_{\bm k}(\eta) = \left\{
\begin{array}{ll}
A_1\epsilon & \quad (\text{long-wavelength})
\vspace{0.5em}
\\
4\pi G \dot{\phi}_0 [c_1\sin(k\eta) + c_2 \cos(k\eta)]
& \quad (\text{short-wavelength})
\end{array}
\right.
  ,
\end{equation}
where $A_1$ and $c_i$ are complex constants.

\subsection{Longitudinal gauge}

Let us take the longitudinal gauge by imposing
\begin{equation}
\beta = E = 0   .
\end{equation}
This in turn gives $Q = 0$.
$\tau_{\mu\nu}$ in this gauge is found to be described solely by the gauge-invariant variable $\Psi$,
\begin{align}
\label{eq:2emg-long1}
\tau_{00} 
& = 
\frac{1}{8\pi G  \big( \mathcal{H}^{\prime} -\mathcal{H}^{2} \big)} 
\Bigg\{ 
-\big\langle (\nabla \Psi^{\prime})^{2} \big\rangle 
-\big\langle (\Delta \Psi)^{2} \big\rangle
-2(K +\mathcal{H}) \big\langle \nabla \Psi^{\prime} \cdot 
\nabla \Psi \big\rangle 
- \Big[ 3(\mathcal{H}^{\prime} - \mathcal{H}^{2}) + K^{2} 
+ a^{2} V_{\phi \phi} \Big] 
\big\langle (\Psi^{\prime})^{2} \big\rangle
\nonumber \\
& \hspace{8em}
+\big( 9\mathcal{H}^{\prime} - 10\mathcal{H}^{2}  
- 2L \big) \big\langle (\nabla \Psi)^{2} \big\rangle    
+ 2\bigg[ 6\mathcal{H}(\mathcal{H}^{\prime} -\mathcal{H}^{2} ) 
- KL + 2(\mathcal{H}^{\prime} - \mathcal{H}^{2})  a^{2} \frac{V_{\phi}}{\phi_{0}^{\prime}} 
- a^{2}\mathcal{H} V_{\phi\phi} \bigg] 
 \big\langle \Psi \Psi^{\prime} \big\rangle    
 \nonumber \\
&\hspace{8em}
- \bigg[ L^{2} - 4 \mathcal{H}  (\mathcal{H}^{\prime} - \mathcal{H}^{2})
  a^{2} \frac{V_{\phi}}{\phi_{0}^{\prime}} 
+ a^{2}\mathcal{H}^{2} V_{\phi\phi}  \Bigg\}
\big\langle \Psi^{2} \big\rangle 
\Bigg]     
  ,
\\
\label{eq:2emg-long2}
\tau_{ij} 
& = 
\frac{\delta_{ij}}{8\pi G  (\mathcal{H}^{\prime} -\mathcal{H}^{2}) } 
\Bigg\{ \frac{1}{3}\big\langle (\nabla \Psi^{\prime})^{2} \big\rangle 
-  \big\langle (\Delta \Psi)^{2} \big\rangle
+ 2\bigg(\frac{\mathcal{H}}{3} - K \bigg)
\big\langle \nabla \Psi^{\prime} \cdot \nabla \Psi \big\rangle 
+ \Big( \mathcal{H}^{\prime} - \mathcal{H}^{2} - K^{2} + a^{2} V_{\phi \phi}  \Big) 
\big\langle (\Psi^{\prime})^{2} \big\rangle
\nonumber \\
& \hspace{8em}
+\bigg[ \frac{1}{3} \big( 11\mathcal{H}^{\prime} - 10\mathcal{H}^{2} \big) 
- 2L \bigg] \big\langle (\nabla \Psi)^{2} \big\rangle    
+ 2 \Big[  \mathcal{H} a^{2} V_{\phi\phi}  - KL
+ 2 (\mathcal{H}^{\prime} -\mathcal{H}^{2} ) ( K + 2 \mathcal{H} )  \Big] 
 \big\langle \Psi \Psi^{\prime} \big\rangle    
 \nonumber \\
& \hspace{8em}
+ \Big[ 4 ( \mathcal{H}^{\prime} - \mathcal{H}^{2}) 
(\mathcal{H}^{\prime} +2\mathcal{H}^{2} + L) 
- L^{2}  + \mathcal{H}^{2} a^{2} V_{\phi\phi}  \Big]
\big\langle \Psi^{2} \big\rangle 
\Bigg\}     
  .
\end{align}

\subsubsection{Long-wavelength limit}

In the long-wavelength limit $k \ll \calH$, 
we find $\tau_{\mu\nu}$ in the longitudinal gauge as
\begin{align}
\tau_{00} 
& \approx 
\frac{|A_1|^2 k^2}{8\pi G}\left[ 2\epsilon +3\frac{{\cal  H}^2}{k^2}(-3\epsilon^2+\epsilon\delta) \right]
  ,
\\
\tau_{ij} 
& \approx 
\frac{|A_1|^2 k^2}{8\pi G}\left[ -\frac{2}{3}\epsilon +3\frac{{\cal  H}^2}{k^2}(3\epsilon^2-\epsilon\delta) \right]
  ,
\end{align}
We note that the long-wavelength limit can be divided into two sub-limits, depending on which term inside the square brackets dominates; (i) the ultra long-wavelength limit for which the second term dominates so that ${\cal H}/k \gtrsim 1/\sqrt{\epsilon}$, and (ii) the infra long-wavelength limit for which the first term dominates so that ${\cal H}/k \lesssim 1/\sqrt{\epsilon}$.

\begin{enumerate}[label=(\roman*)]
\item Ultra long-wavelength limit: In this limit, $k \lesssim \sqrt{\epsilon}\calH$ so that 
the second term, i.e., ${\cal O}(\epsilon^2)$ term in the 2EMT is dominant.
The value of the equation-of-state parameter $w$ in this limit is given by
\begin{align}
w & \equiv  \frac{\mathfrak{p}}{\varrho} \approx -1   ,
\\
\varrho & \approx 
-\frac{3H^2|A_1|^2(3\epsilon^2-\epsilon\delta)}{8\pi G}   ,
\end{align} 
where the the effective energy density and pressure of 2EMT are given by
$\varrho = \tau_{00}/a^2$ and $\mathfrak{p}= \tau_{ii}/a^2$.
Note that this result agrees with that in Ref.~\cite{Abramo:1997hu} 
in which the authors showed that $w \approx -1$ with $\varrho <0$.

\item Infra long-wavelength limit: In this limit, $k \gtrsim \sqrt{\epsilon}\calH$ so that 
the first term, i.e., ${\cal O}(\epsilon)$ term in 2EMT is dominant.
The value of $w$ in this limit is given by
\begin{align}
w & \approx -\frac{1}{3}   ,
\\
\varrho & \approx \frac{k^2 |A_1|^2\epsilon}{4\pi Ga^2} >0   .
\end{align} 
\end{enumerate}

It is interesting to note that the scale dividing the ultra and infra long-wavelength limits coincides with the so-called decoupling limit, where we can neglect the perturbations in the metric and may only consider those in the matter sector. That is, the gravitational background is considered to be completely classical, and the Goldstone boson picture for the curvature perturbation becomes manifest~\cite{Cheung:2007st}
-- only the scalar field is quantum mechanical in the classical gravitational background. 
Interestingly, in such a limit the equation of state of 2EMT is 
$w \approx-1/3$ in {\it all gauge choices} we examine in this section.

\subsubsection{Short-wavelength limit}

In the short-wavelength limit $k \gg \calH$, 
we find  $\tau_{\mu\nu}$ as
\begin{align}
\tau_{00} 
& \approx 
\left( |c_1|^2+|c_2|^2 \right) \frac{k^4}{4a^2} 
\left[ 6 + \frac{\mathcal{H}^2}{k^2} (14\epsilon -\delta) \right],
\\
\tau_{ij} 
& \approx 
\left( |c_1|^2+|c_2|^2 \right) 
\frac{k^4}{4a^2} 
\left[ \frac{10}{3} + \frac{\mathcal{H}^2}{k^2} \bigg( \frac{4\epsilon}{3} -\frac{5\delta}{3} \bigg) \right].
\end{align}
The value of $w$ in this limit is given by
\begin{align}
w & \approx \frac{5}{9}   , 
\\
\varrho & \approx \frac{3k^4 ( |c_1|^2+|c_2|^2 )}{2a^4}   .
\end{align}

\subsection{Spatially flat gauge}

Let us take the spatially flat gauge,
\begin{equation}
\psi = E = 0,
\end{equation}
so that the spatial metric is unperturbed. It is equivalent to set
\begin{equation}
Q = \frac{\Psi}{\calH}.
\end{equation}
$\tau_{\mu\nu}$ in this gauge is then found to be described solely by $\Psi$,
\begin{align}
\label{eq:2emg-flat1}
\tau_{00} 
& = \frac{1}{8\pi G (\mathcal{H}^{\prime} -\mathcal{H}^{2}) } \Bigg\{
-\big\langle (\nabla \Psi^{\prime})^{2} \big\rangle 
-\big\langle (\Delta \Psi)^{2} \big\rangle
+2(L-M) \big\langle \nabla \Psi^{\prime} \cdot 
\nabla \Psi \big\rangle 
\nonumber \\
& \hspace{1in}
- \bigg[ \bigg( K +  \frac{3 (\mathcal{H}^{\prime} -\mathcal{H}^{2})}{ \mathcal{H} } \bigg)^{2}
- \frac{4}{\mathcal{H}^{2}}  (\mathcal{H}^{\prime} -\mathcal{H}^{2})
 (2\mathcal{H}^{\prime} + \mathcal{H}^{2}) + a^{2} V_{\phi \phi} 
 \bigg] 
\big\langle (\Psi^{\prime})^{2} \big\rangle       
\nonumber \\
&\hspace{1in}
- \frac{1} {\mathcal{H}^{2}}  \Big[ (\mathcal{H}^{\prime} -2 \mathcal{H}^{2}) 
(\mathcal{H}^{\prime} - 2 \mathcal{H}^{2} - 2L)
 + (\mathcal{H}^{\prime} -4\mathcal{H}^{2})
(\mathcal{H}^{\prime} -\mathcal{H}^{2})  \Big]  \big\langle (\nabla \Psi)^{2} \big\rangle  
\nonumber \\
&\hspace{1in}
- \bigg[ \frac{2L(\mathcal{H}^{\prime} -\mathcal{H}^{2})}{ \mathcal{H}^{2}}
\bigg( 3K - \frac{\mathcal{H}^{\prime} + 11 \mathcal{H}^{2}}{\mathcal{H}} \bigg)
- \frac{2(\mathcal{H}^{\prime} -2 \mathcal{H}^{2}) }{\mathcal{H}}
 a^{2} V_{\phi \phi}  \bigg] 
\big\langle \Psi^{\prime} \Psi \big\rangle       
\nonumber \\
&\hspace{1in}
- \frac{(\mathcal{H}^{\prime} - 2 \mathcal{H}^{2})^{2}}{\mathcal{H}^{2}} 
\bigg[ \bigg( L -   \frac{2 (\mathcal{H}^{\prime} - \mathcal{H}^{2}) }{\mathcal{H}} \bigg)^{2}
+ 12  (\mathcal{H}^{\prime} - \mathcal{H}^{2})
+ a^{2} V_{\phi \phi}  \bigg] 
\big\langle \Psi^{2} \big\rangle   \Bigg\}      
  ,
\\
\label{eq:2emg-flat2}
\tau_{ij} 
& = 
\frac{\delta_{ij}}{8\pi G {\mathcal{H}^{2}} (\mathcal{H}^{\prime} - \mathcal{H}^{2}) } 
\Bigg\{ 
\frac{1}{3} (2 \mathcal{H}^{\prime} - \mathcal{H}^{2})
 \big\langle (\nabla \Psi^{\prime})^{2} \big\rangle 
- \frac{1}{3}(\mathcal{H}^{\prime} + 2 \mathcal{H}^{2}) 
 \big\langle (\Delta \Psi)^{2} \big\rangle     
 \nonumber \\
&
\hspace{9em}
- \frac{2}{3 \mathcal{H} }
\Big[ 4  \mathcal{H}^{\prime 2} + 3  \mathcal{H}^{\prime} \mathcal{H}^{2}
- 8 \mathcal{H}^{4} + (2 \mathcal{H}^{\prime} + \mathcal{H}^{2}) \mathcal{H} K
\Big] \big\langle \nabla \Psi^{\prime} \cdot  \nabla \Psi \big\rangle       
\nonumber \\
&
\hspace{9em}
-\Big[ 
3 ( \mathcal{H}^{\prime} -\mathcal{H}^{2} ) ( 3\mathcal{H}^{\prime} -7 \mathcal{H}^{2} )
+ 14 \mathcal{H} ( \mathcal{H}^{\prime} -\mathcal{H}^{2} ) K 
+ \mathcal{H}^{2} K^{2} -   \mathcal{H}^{2} a^{2} V_{\phi \phi}  
 \Big] 
\big\langle (\Psi^{\prime})^{2} \big\rangle       
\nonumber \\
&
\hspace{9em}
+ \frac{1 } {3 \mathcal{H}^{2}} ( \mathcal{H}^{\prime} -2 \mathcal{H}^{2} )  \Big[ 
2 \mathcal{H}^{\prime 2 } + 10 \mathcal{H}^{\prime} \mathcal{H}^{2}
-13  \mathcal{H}^{4} + 2 (2 \mathcal{H}^{\prime} + \mathcal{H}^{2}) L
  \Big]  \big\langle (\nabla \Psi)^{2} \big\rangle  
  \nonumber \\
&
\hspace{9em}
+\bigg[ \frac{2}{\mathcal{H}} (\mathcal{H}^{\prime} -2 \mathcal{H}^{2} )
\Big( L^{2} + 12 (\mathcal{H}^{\prime} - \mathcal{H}^{2}) L
- 4 (\mathcal{H}^{\prime} + 2 \mathcal{H}^{2})(\mathcal{H}^{\prime} - \mathcal{H}^{2})  \Big)
- 2 \mathcal{H} (\mathcal{H}^{\prime} - \mathcal{H}^{2} )  
a^{2} V_{\phi \phi}   \bigg]       
\big\langle \Psi^{\prime} \Psi \big\rangle 
\nonumber \\
&
\hspace{9em}
- \frac{ (\mathcal{H}^{\prime} -2 \mathcal{H}^{2} )}{\mathcal{H}^{2}}
\Big[ L^{2} + 12  (\mathcal{H}^{\prime} - \mathcal{H}^{2}) L
- \mathcal{H}^{2} a^{2} V_{\phi \phi} 
- 4 (\mathcal{H}^{\prime} + 2 \mathcal{H}^{2} )(\mathcal{H}^{\prime} - \mathcal{H}^{2} )
\Big] 
\big\langle \Psi^{2} \big\rangle   \ 
\Bigg\}
  .
\end{align}

\subsubsection{Long-wavelength limit}

In the long-wavelength limit $k \ll \calH$, 
we find $\tau_{\mu\nu}$ in the spatially flat gauge as
\begin{align}
\tau_{00} 
& \approx 
\frac{|A_1|^2 k^2}{8\pi G} \left[ \epsilon +3\frac{{\cal  H}^2}{k^2}(-3\epsilon^2+\epsilon\delta) \right]
  ,
\\
\tau_{ij} 
& \approx 
\frac{|A_1|^2 k^2}{8\pi G} \left[ -\frac{1}{3}\epsilon +3\frac{{\cal  H}^2}{k^2}(3\epsilon^2-\epsilon\delta) \right]
  ,
\end{align}
Again, the long-wavelength limit is divided into the ultra and infra long-wavelength limits.

\begin{enumerate}[label=(\roman*)]

\item Ultra long-wavelength limit: In this limit, the result is exactly the same with that of the longitudinal gauge.

\item Infra long-wavelength limit: In this limit, the value of $w$ in this limit is the same as that in the longitudinal gauge, but the energy density is smaller by factor 2,
\begin{align}
w & \approx -\frac{1}{3}   ,
\\
\varrho & \approx \frac{k^2 |A_1|^2\epsilon}{8\pi Ga^2}   .
\end{align} 

\end{enumerate}

\subsubsection{Short-wavelength limit}

In the short-wavelength limit $k \gg \calH$, we find
\begin{align}
\tau_{00} 
& \approx 
\left( |c_1|^2+|c_2|^2 \right)  \frac{k^4}{4a^2} 
\left[ 2 + \frac{\mathcal{H}^2}{k^2} (-4\epsilon +\delta) \right]
  ,
\\
\tau_{ij} 
& \approx 
\left( |c_1|^2+|c_2|^2 \right) \frac{k^4}{4a^2}  
\left[ \frac{2}{3} + \frac{\mathcal{H}^2}{k^2} \left( 2\epsilon - \frac{7\delta}{3} \right) \right]
  .
\end{align}
And the value of $w$ in this limit is given by
\begin{align}
w & \approx \frac{1}{3}   ,
\\
\varrho & \approx \frac{k^4 ( |c_1|^2+|c_2|^2 )}{2a^4}   .
\end{align}

\subsection{Comoving gauge}

Let us impose the comoving gauge condition in the scalar mode,
\begin{equation}
\delta\phi = E = 0   ,
\end{equation}
for which the gauge-invariant variable $\Psi$ and the gauge-dependent variable $Q$ are given by
\begin{align}
\Psi & = \psi + \mathcal{H} \beta   ,
\\
\label{eq:Qcomoving}
Q & =  \frac{\Psi^{\prime} + \mathcal{H} \Psi}{\mathcal{H}^{\prime} - \mathcal{H}^{2}}   . 
\end{align}
$\tau_{\mu\nu}$ in this gauge is found to be
\begin{align}
\label{eq:2emt-com1}
\tau_{00} 
& = \frac{1}{8\pi G (\mathcal{H}^{\prime} -\mathcal{H}^{2})^2 } \Bigg[
-{2 \mathcal{H}} \big\langle   \Delta \Psi^{\prime} \Delta \Psi       \big\rangle
- \big( 4\mathcal{H}^{\prime} - 3 \mathcal{H}^{2} \big)
\big\langle     \big( \Delta \Psi  \big)^{2}       \big\rangle  
\nonumber \\
& \hspace{8.5em}
+ \frac{F_{1}}{ \mathcal{H}^{\prime} - \mathcal{H}^{2} }
\big\langle     \big( \nabla \Psi^{\prime}  \big)^{2}     \big\rangle
+ \frac{2 F_{2}}{ \mathcal{H}^{\prime} - \mathcal{H}^{2} }
\big\langle      \nabla \Psi^{\prime} \cdot \nabla \Psi      \big\rangle
+ \frac{F_{3}}{ \mathcal{H}^{\prime} - \mathcal{H}^{2} }
\big\langle  \big(  \nabla \Psi \big)^{2}    \big\rangle
\nonumber \\
& \hspace{8.5em}
- \frac{F_{4}}{ \big( \mathcal{H}^{\prime} - \mathcal{H}^{2} \big)^{2}}
\big\langle  \big(  \Psi^{\prime} \big)^{2}    \big\rangle
- \frac{2 F_{5}}{ \big( \mathcal{H}^{\prime} - \mathcal{H}^{2} \big)^{2}}
\big\langle  \Psi^{\prime} \Psi   \big\rangle
- \frac{F_{6}}{ \big( \mathcal{H}^{\prime} - \mathcal{H}^{2} \big)^{2}}
\big\langle  \Psi^{2} \big\rangle   
\Bigg]      
  ,
\\
\label{eq:2emt-com2}
\tau_{ij} 
& = 
\frac{\delta_{ij}}{8\pi G (\mathcal{H}^{\prime} - \mathcal{H}^{2})^2 } 
\Bigg[ 
- \frac{2}{3} \big\langle  \big( \Delta \Psi^{\prime} \big)^{2}  \big\rangle 
- \frac{2}{3} \big\langle   \Delta^{2} \Psi \Delta \Psi    \big\rangle 
 + \frac{2P_{1}}{3 \big( \mathcal{H}^{\prime} - \mathcal{H}^{2} \big)}
\big\langle \Delta \Psi^{\prime} \Delta \Psi     \big\rangle 
- \frac{P_{2}}{3 \big( \mathcal{H}^{\prime} - \mathcal{H}^{2} \big)}
\big\langle \big( \Delta \Psi \big)^{2} \big\rangle 
\nonumber \\
& \hspace{8.5em}
+ \frac{P_{3}}{ 3 \big( \mathcal{H}^{\prime} - \mathcal{H}^{2} \big)^{2}}
\big\langle     \big( \nabla \Psi^{\prime}  \big)^{2}     \big\rangle
- \frac{2 P_{4}}{ 3 \big( \mathcal{H}^{\prime} - \mathcal{H}^{2} \big)^{2}}
\big\langle     \nabla \Psi^{\prime}  \cdot \nabla \Psi    \big\rangle
+ \frac{P_{5}}{ 3 \big( \mathcal{H}^{\prime} - \mathcal{H}^{2} \big)^{2}}
\big\langle     \big( \nabla \Psi  \big)^{2}     \big\rangle
\nonumber \\
& \hspace{8.5em}
- \frac{P_{6}}{ 3 \big( \mathcal{H}^{\prime} - \mathcal{H}^{2} \big)^{3}}
\big\langle     \big( \Psi^{\prime}  \big)^{2}     \big\rangle 
+ \frac{2 P_{7}}{ \big( \mathcal{H}^{\prime} - \mathcal{H}^{2} \big)^{3}}
\big\langle      \Psi^{\prime}  \Psi     \big\rangle
- \frac{P_{8}}{ 3 \big( \mathcal{H}^{\prime} - \mathcal{H}^{2} \big)^{3}}
\big\langle      \Psi^{2}     \big\rangle    
\Bigg]
  .
\end{align} 
The time-dependent coefficients $F_i$ and $P_i$ are given in Appendix~\ref{appsec:com-coeff}.

\subsubsection{Long-wavelength limit}

In the long-wavelength limit $k \ll \calH$, we find in the comoving gauge
\begin{align}
\tau_{00} 
& \approx 
\frac{|A_1|^2 {\cal  H}^2}{8\pi G} \Big[ 9 +(49\epsilon-36\delta) + 4(16\epsilon^2 -25\epsilon\delta+9\delta^2) \Big],
\\
\tau_{ij} 
& \approx 
\frac{|A_1|^2 {\cal  H}^2}{8\pi G} \Big[ -3 +(5\epsilon+12\delta) + 4(13\epsilon^2 -8\epsilon\delta -3\delta^2) \Big].
\end{align}
In the comoving gauge, unlike the other gauges we have examined previously, the long-wavelength limit is not sub-divided into ultra and infra long-wavelength limits. The value of $w$ in the long-wavelength limit is the same as the infra long-wavelength limit of the previous gauge choices, but with a different energy density,
\begin{align}
w & \approx -\frac{1}{3}   ,
\\
\varrho & \approx \frac{9 {\cal H}^2 |A_1|^2}{8\pi Ga^2}
=  \frac{9 H^2 |A_1|^2}{8\pi G}   . 
\end{align}

\subsubsection{Short-wavelength limit}

In the short-wavelength limit $k \gg \calH$, we find
\begin{align}
\tau_{00} 
& \approx 
\left( |c_1|^2+|c_2|^2 \right) \frac{k^4}{4a^2} 
\left[ -\frac{1}{\epsilon} + \left( 3 -\frac{2\delta}{\epsilon} \right)   
+ \frac{\mathcal{H}^2}{k^2} (5\epsilon +2\delta) \right],
\\
\tau_{ij} 
& \approx 
-\left( |c_1|^2+|c_2|^2 \right)\frac{k^4}{4a^2} 
\left[ \frac{11}{3\epsilon} +\left( 1+\frac{2\delta}{3\epsilon}\right) + \frac{2\delta^2}{\epsilon} \right].
\end{align}
The value of $w$ in this limit is given by
\begin{align}
w & \approx \frac{11}{3}   ,
\\
\varrho & \approx -\frac{k^4 ( |c_1|^2+|c_2|^2 )}{4a^4\epsilon}   .
\end{align}

\section{Conclusions}
\label{sec:conclusion}

We considered the cosmological perturbations in the universe dominated by a scalar field.
Introducing the scalar modes of the metric perturbations and the matter field perturbation,
we derived the Einstein's equation up to the second order, and constructed 
2EMT, which is composed of the quadratic terms of the linear perturbations, 
and is responsible the so-called gravitational back-reaction.

Like the first order, the second-order Einstein's equation may be put into a gauge invariant form
after adding terms on both sides of the equation.
However, 2EMT which are the collection of only quadratic terms of the equation
is not guaranteed of its gauge invariance.
Some works have been done in this context, e.g., 
in Refs.~\citep{Geshnizjani:2002wp,Brandenberger:2002sk,Geshnizjani:2003cn,Martineau:2005aa,Martineau:2005zu}.
However, the gauge invariance of 2EMT was doubted in 
Refs.~\cite{Unruh:1998ic,Ishibashi:2005sj,Green:2010qy}.

In this work, we derived 2EMT directly and considered its gauge dependence.
As we observed in the Friedmann universe driven by perfect fluid \cite{Cho:2020zbh},
2EMT in the inflation period driven by a scalar field
also exhibits gauge dependence; 2EMT does not depend only on the gauge invariant variable, but also on the gauge dependent variables.

Although 2EMT is gauge dependent, we investigated it in certain wavelength limits
under different gauge conditions in order to examine if its behaviour converges.
We examined in three conditions (longitudinal, spatially flat, and comoving gauges) in the slow-roll regime of inflation.
The result showed that 2EMT behaves as an effective fluid of $w=-1/3$
in all gauges in the long-wavelength limit.
Specifically, it was observed in the {\it infra} long-wavelength limit, 
$\sqrt{\epsilon}{\cal H} \lesssim k \ll {\cal H}$.
This behaviour was not observed in the Friedmann universe dominated by fluid in Ref.~\cite{Cho:2020zbh}.

\subsection*{Acknowledgements}

This work is supported in parts by the grants from the National Research Foundation
funded by the Korean government, 
2020R1A2C1013266 (I.C.), 2019R1A2C2085023 (J.G.), and 2021R1F1A1063308 (S.H.O.). 
J.G. also acknowledges the Korea-Japan Basic Scientific Cooperation Program supported 
by the National Research Foundation of Korea and the Japan Society 
for the Promotion of Science (2020K2A9A2A08000097). 
J.G. is supported in part by the Ewha Womans University Research Grant of 
2020 (1-2020-1630-001-1), 2021 (1-2021-1227-001-1) and 2022 (1-2022-0606-001-1).
J.G. is grateful to the Asia Pacific Center for Theoretical Physics for hospitality 
while this work was under progress.

\appendix

\section{Time-dependent coefficients in the comoving gauge}
\label{appsec:com-coeff}

Here, we list the time-dependent coefficients $F_i$ and $P_i$ of the 2EMT in the comoving gauge
in Eqs.~\eqref{eq:2emt-com1} and \eqref{eq:2emt-com2}.
\begin{align}
F_{1} = & 2 \mathcal{H} \mathcal{H}^{ \prime \prime}+ (\mathcal{H}^{\prime})^{2}
+ 4 \mathcal{H} (K- 2 \mathcal{H}) \mathcal{H}^{\prime} 
- \mathcal{H}^{3} (4K - 3 \mathcal{H}) ,
\\
F_{2} = & - ( 2\mathcal{H}^{\prime} -\mathcal{H}^{2}) \mathcal{H}^{\prime \prime}
- (4 K- 5\mathcal{H})( \mathcal{H}^{\prime})^{2}
+ 2 \mathcal{H} ( 2K\mathcal{H} + L  - 3 \mathcal{H}^{2}) \mathcal{H}^{\prime}
- \mathcal{H}^{3} ( 2L-3 \mathcal{H}^{2}) ,
\\
F_{3} = & - 4 \mathcal{H} \mathcal{H}^{\prime} \mathcal{H}^{\prime\prime}
+ 13 (\mathcal{H}^{\prime})^{3} - \left[ 8L - 2 \left( K - \frac{a^{2} V_\phi}{\phi_{0}^{\prime}} \right)
 \mathcal{H} + 12 \mathcal{H}^{2}  \right] (\mathcal{H}^{\prime})^{2}  
 \nonumber \\
 &
 - \mathcal{H}^{2} \left( 4 \mathcal{H} K - 8L + \mathcal{H}^{2}
 - 4 \mathcal{H}   \frac{a^{2} V_\phi}{\phi_{0}^{\prime}}    \right)  \mathcal{H}^{\prime}
 + 2  \mathcal{H}^{5} \left( K + 4 \mathcal{H} - \frac{a^{2} V_\phi}{\phi_{0}^{\prime}}  \right) ,
\\
 F_{4} = & ( \mathcal{H}^{\prime} + 2 \mathcal{H}^{2} ) (\mathcal{H}^{\prime\prime})^{2}
 -4 \big[ 2 \mathcal{H} (\mathcal{H}^{\prime})^{2}  - K (\mathcal{H}^{\prime})^{2}
 - K \mathcal{H}^{2} \mathcal{H}^{\prime} + \mathcal{H}^{4} (2K + \mathcal{H})
 \big] \mathcal{H}^{\prime \prime} 
 \nonumber \\
 &
 + \left[  5K^{2} - 16 K \mathcal{H} + 7 \mathcal{H}^{2} 
 - 2K a^{2} V^{\prime}(\phi_{0})/\phi_{0}^{\prime}
 + \left( \frac{a^{2} V_\phi}{\phi_{0}^{\prime}} \right)^{2}
   \right] (\mathcal{H}^{\prime})^{3}  
\nonumber \\
&
 - \left[ 3K^{2} -16\mathcal{H}K + 5 \mathcal{H}^{2}  
 - 6K a^{2} V^{\prime}(\phi_{0})/\phi_{0}^{\prime}
 +3 \left( \frac{a^{2} V_\phi}{\phi_{0}^{\prime}} \right)^{2}
 \right] \mathcal{H}^{2} (\mathcal{H}^{\prime})^{2} 
 \nonumber \\
&
 - \left[ 9 K^{2} + 8 \mathcal{H}K -17 \mathcal{H}^{2}  
 + 6K a^{2} V^{\prime}(\phi_{0})/\phi_{0}^{\prime}
 -3  \left( \frac{a^{2} V_\phi}{\phi_{0}^{\prime}} \right)^{2}
 \right] \mathcal{H}^{4} (\mathcal{H}^{\prime}) 
 \nonumber \\
&
 + \left[ 7K^{2} +8\mathcal{H}K -7 \mathcal{H}^{2}  
 + 2 K a^{2} V^{\prime}(\phi_{0})/\phi_{0}^{\prime}
 - \left( \frac{a^{2} V_\phi}{\phi_{0}^{\prime}} \right)^{2}
 \right] \mathcal{H}^{6},
\\
F_{5} = & ( \mathcal{H}^{\prime} + 2 \mathcal{H}^{2} ) \mathcal{H}
(\mathcal{H}^{\prime\prime})^{2}  - \left[ (\mathcal{H}^{\prime})^{3}  
- \left( 2 \mathcal{H} K + L - 14 \mathcal{H}^{2} +\mathcal{H}
\frac{a^{2} V_\phi}{\phi_{0}^{\prime}}     \right) (\mathcal{H}^{\prime})^{2}  
 \right.
 \nonumber \\
 &
\left.
- \left( 2 \mathcal{H} K + 4 L + 15 \mathcal{H}^{2} - 2 \mathcal{H}
\frac{a^{2} V_\phi}{\phi_{0}^{\prime}}    \right) \mathcal{H}^{2}  \mathcal{H}^{\prime}
+ \left( 4 \mathcal{H} K + 5 L + 12 \mathcal{H}^{2} - \mathcal{H}
\frac{a^{2} V_\phi}{\phi_{0}^{\prime}}    \right) \mathcal{H}^{4}  
\right] \mathcal{H}^{\prime \prime}
\nonumber \\
 &
 - 2 (K - 2 \mathcal{H})(\mathcal{H}^{\prime})^{4}
 + \left[ 3KL - 18 \mathcal{H}^{2} K -7 \mathcal{H} L 
 + ( \mathcal{H} K - L - \mathcal{H}^{2} )  \frac{a^{2} V_\phi}{\phi_{0}^{\prime}}
 +  \mathcal{H} \left( \frac{a^{2} V_\phi}{\phi_{0}^{\prime}}  \right)^{2}
   \right] (\mathcal{H}^{\prime})^{3}
 \nonumber \\
 &
 + \left[ - \mathcal{H}^{2} \frac{a^{2} V_\phi}{\phi_{0}^{\prime}} 
 \left(  3\mathcal{H} K - 3 L - \mathcal{H}^{2}  
 + 3 \mathcal{H} \frac{a^{2} V_\phi}{\phi_{0}^{\prime}}    \right)
 + \mathcal{H}^{2} \big( 3KL + 50 \mathcal{H}^{2} K + 7 \mathcal{H} L 
 - 68 \mathcal{H}^{3} \big)
   \right] (\mathcal{H}^{\prime})^{2}
   \nonumber \\
 &
 + \left[  \mathcal{H}^{4} \frac{a^{2} V_\phi}{\phi_{0}^{\prime}}
 \left(  3\mathcal{H} K - 3 L +  \mathcal{H}^{2}  
 + 3 \mathcal{H} \frac{a^{2} V_\phi}{\phi_{0}^{\prime}}    \right)
  - 5 \mathcal{H}^{4} \big( 3KL + 10 \mathcal{H}^{2} K +  \mathcal{H} L 
 - 14 \mathcal{H}^{3} \big)
   \right] \mathcal{H}^{\prime}
   \nonumber \\
 &
 + \left[  - \frac{a^{2} V_\phi}{\phi_{0}^{\prime}} 
 \left(  \mathcal{H} K -  L +  \mathcal{H}^{2}  
 +  \mathcal{H} \frac{a^{2} V_\phi}{\phi_{0}^{\prime}}    \right)
  + 9KL + 20 \mathcal{H}^{2} K + 5 \mathcal{H} L 
 - 20 \mathcal{H}^{3}    \right] \mathcal{H}^{6} ,
 \\
 F_{6} = & ( \mathcal{H}^{\prime} + 2 \mathcal{H}^{2} ) \mathcal{H}^{2}
(\mathcal{H}^{\prime\prime})^{2} 
\nonumber \\
&
+ 2 \mathcal{H} \left[ 
- (\mathcal{H}^{\prime})^{3}  + \left( L - 10 \mathcal{H}^{2} 
+ \mathcal{H}   \frac{a^{2} V_\phi}{\phi_{0}^{\prime}}  \right) 
(\mathcal{H}^{\prime})^{2} 
+ \left( 4 L + 15 \mathcal{H}^{2} 
-2  \mathcal{H}   \frac{a^{2} V_\phi}{\phi_{0}^{\prime}}  \right) 
 \mathcal{H}^{2}  \mathcal{H}^{\prime}  
 - \mathcal{H}^{4} \left( 5L + 10\mathcal{H}^{2} 
- \mathcal{H}  \frac{a^{2} V_\phi}{\phi_{0}^{\prime}}  \right) 
\right] \mathcal{H}^{\prime \prime} 
\nonumber \\
&
+ (\mathcal{H}^{\prime})^{5}
- \left( 2L - 33\mathcal{H}^{2} + 2\mathcal{H}  \frac{a^{2} V_\phi}{\phi_{0}^{\prime}}
 \right) (\mathcal{H}^{\prime})^{4}
 + \left[ \mathcal{H} \frac{a^{2} V_\phi}{\phi_{0}^{\prime}} 
 \left( 2L + 6  \mathcal{H}^{2} + \mathcal{H}  \frac{a^{2} V_\phi}{\phi_{0}^{\prime}}  \right)
 + L^{2} - \mathcal{H}^{2} (42 L + 65 \mathcal{H}^{2})
 \right] (\mathcal{H}^{\prime})^{3}
 \nonumber \\
 &
 +  \left[ - \mathcal{H} \frac{a^{2} V_\phi}{\phi_{0}^{\prime}}
 \left( 6 L + 10  \mathcal{H}^{2} + 3 \mathcal{H}  \frac{a^{2} V_\phi}{\phi_{0}^{\prime}}  \right)
  +  9 L^{2} +110 L  \mathcal{H}^{2} + 39 \mathcal{H}^{4} 
\right] \mathcal{H}^{2} (\mathcal{H}^{\prime})^{2} 
\nonumber \\
&
 +  \left[ \mathcal{H} \frac{a^{2} V_\phi}{\phi_{0}^{\prime}}
 \left( 6 L + 10  \mathcal{H}^{2} + 3 \mathcal{H}  \frac{a^{2} V_\phi}{\phi_{0}^{\prime}}  \right)
 -21 L^{2} -110 L  \mathcal{H}^{2} + 8 \mathcal{H}^{4} 
\right] \mathcal{H}^{4} \mathcal{H}^{\prime}
\nonumber \\
&
 + \left[  - \mathcal{H} \frac{a^{2} V_\phi}{\phi_{0}^{\prime}}
 \left( 2 L + 4  \mathcal{H}^{2} +  \mathcal{H}  \frac{a^{2} V_\phi}{\phi_{0}^{\prime}}  \right)
 + 11L^{2} + 44 \mathcal{H}^{2} L - 4 \mathcal{H}^{4}
\right] \mathcal{H}^{6} ,
\\
P_{1} = & 4 \mathcal{H}^{\prime \prime} + (6K - 11 \mathcal{H}) \mathcal{H}^{\prime}
- 3 \mathcal{H}^{2} (2K- \mathcal{H}),
\\
P_{2} = & 4  \mathcal{H}  \mathcal{H}^{\prime \prime} + 4 ( \mathcal{H}^{\prime})^{2} 
+ (8 \mathcal{H} K -8L - 15  \mathcal{H}^{2} )  \mathcal{H}^{\prime}
-  \mathcal{H}^{2} (8 \mathcal{H} K -8L - 3  \mathcal{H}^{2})
\\
P_{3} = & -2 (\mathcal{H}^{\prime} - \mathcal{H}^{2}) \mathcal{H}^{\prime \prime\prime}
+ 6 ( \mathcal{H}^{\prime\prime})^{2} + 2 \big[ ( 8K -13 \mathcal{H} ) \mathcal{H}^{\prime}
- \mathcal{H}^{2} ( 8K -\mathcal{H} )   \big] \mathcal{H}^{\prime\prime}
+ 9 (\mathcal{H}^{\prime})^{3} 
\nonumber \\
&
+ \big[ 8K( 2K - 5 \mathcal{H} ) - 4L + 27 \mathcal{H}^{2}  \big] (\mathcal{H}^{\prime})^{2}
- \big[ 16K( 2K - 3 \mathcal{H} ) - 8 L + 17 \mathcal{H}^{2}  \big]  \mathcal{H}^{2} \mathcal{H}^{\prime}
\nonumber \\
&
+ \big[ 8K( 2K -  \mathcal{H} ) - 4 L + 5 \mathcal{H}^{2}  \big] \mathcal{H}^{4},
\\
P_{4} = & - \mathcal{H} (\mathcal{H}^{\prime} - \mathcal{H}^{2}) \mathcal{H}^{\prime \prime\prime}
+ 6 \mathcal{H} (\mathcal{H}^{\prime \prime})^{2}
- \big[ 3 (\mathcal{H}^{\prime})^{2} - ( 22 \mathcal{H} K - 8 L - 19 \mathcal{H}^{2}  ) \mathcal{H}^{\prime}
+ 2 \mathcal{H}^{2} (11\mathcal{H} K - 4L + \mathcal{H}^{2} )  \big] \mathcal{H}^{\prime \prime}
\nonumber \\
&
- (8K - 29\mathcal{H}) (\mathcal{H}^{\prime})^{3} 
+ \big[  4K ( 5K\mathcal{H} -3 L - 7 \mathcal{H}^{2}  ) 
+ \mathcal{H} (  16 L -69 \mathcal{H}^{2}  )  \big] (\mathcal{H}^{\prime})^{2} 
\nonumber \\
&
- \mathcal{H}^{2} \big[ 4 \mathcal{H} K (10K-9\mathcal{H}) - 8L (3K-2\mathcal{H})
- 107 \mathcal{H}^{3}  \big]  \mathcal{H}^{\prime}
+ \mathcal{H}^{4} \big[ 4K(5\mathcal{H}K - 3L) -43 \mathcal{H}^{3}  \big],
\\
P_{5} = & 4 \mathcal{H}^{2} (\mathcal{H}^{\prime} - \mathcal{H}^{2}) 
\mathcal{H}^{\prime \prime\prime} - 18 \mathcal{H}^{2} (\mathcal{H}^{\prime\prime})^{2}
+ \mathcal{H} \big[  60  (\mathcal{H}^{\prime})^{2} 
- 4 ( 3 \mathcal{H}K + 8L + 11 \mathcal{H}^{2}   )  \mathcal{H}^{\prime}
    \nonumber \\
&
+ 4 \mathcal{H}^{2} ( 3 \mathcal{H} K + 8 L + 14 \mathcal{H}^{2}) \big] \mathcal{H}^{\prime \prime} 
- 23 (\mathcal{H}^{\prime})^{4}
+ (60 \mathcal{H} K + 4 L -131 \mathcal{H}^{2}  )  (\mathcal{H}^{\prime})^{3} 
 \nonumber \\
&
- \big[  4 \mathcal{H} K ( 10L + 39 \mathcal{H}^{2} ) - 4 L ( 2L + 17\mathcal{H}^{2} )
- 333 \mathcal{H}^{4}    \big]  (\mathcal{H}^{\prime})^{2} 
 \nonumber \\
&
+  \mathcal{H}^{2} \big[ 4 \mathcal{H} K ( 20L + 39 \mathcal{H}^{2} ) - 4 L ( 4L + 21\mathcal{H}^{2} )
- 357 \mathcal{H}^{4} \big] \mathcal{H}^{\prime}
 \nonumber \\
&
- 2 \mathcal{H}^{4} \big[ 10 \mathcal{H} K ( 2L + 3 \mathcal{H}^{2} ) - 2 L ( 2L + 3\mathcal{H}^{2} )
- 53 \mathcal{H}^{4} \big] ,
%
%
%
%
\\
P_{6} = & - 6 \mathcal{H} (\mathcal{H}^{\prime} - \mathcal{H}^{2}) 
\big[  \mathcal{H}^{\prime \prime} + 2 (K-\mathcal{H}) \mathcal{H}^{\prime}
- 2 \mathcal{H}^{2} K  \big] \mathcal{H}^{\prime \prime\prime}
+ 12\mathcal{H} (\mathcal{H}^{\prime \prime})^{3} 
- \big[ 9 (\mathcal{H}^{\prime})^{2} - \mathcal{H} (48 K - 57 \mathcal{H} ) \mathcal{H}^{\prime}
+ 6  \mathcal{H}^{3} (8K+ \mathcal{H})  \big] (\mathcal{H}^{\prime \prime})^{2} 
\nonumber \\
&
+ 12 \big[ - (3K -7 \mathcal{H} ) (\mathcal{H}^{\prime})^{3} 
+  \mathcal{H} ( 6K^{2} - 9 \mathcal{H}K - L - 3 \mathcal{H}^{2} )  (\mathcal{H}^{\prime})^{2}
\nonumber \\
&
\qquad
-  \mathcal{H}^{3} (12K^{2} - 11 \mathcal{H} K - 2L - 13\mathcal{H}^{2}) \mathcal{H}^{\prime}
+  \mathcal{H}^{5} (6K^{2} + \mathcal{H} K - L - 5 \mathcal{H}^{2} )
 \big] \mathcal{H}^{\prime \prime}
 \nonumber \\
&
 + 3 \left[  -13 K^{2} + 62 \mathcal{H} K - 49 \mathcal{H}^{2} 
 + \left( \frac{a^{2} V_\phi}{\phi_{0}^{\prime}} \right)^{2}      \right] (\mathcal{H}^{\prime})^{4}
 \nonumber \\
&
 + 12 \mathcal{H} \left[ 4K^{3} -2 \mathcal{H} K^{2} - 32  \mathcal{H}^{2} K - 2KL  + 2 \mathcal{H} L
  + 29 \mathcal{H}^{3}  - \mathcal{H} \left( \frac{a^{2} V_\phi}{\phi_{0}^{\prime}} \right)^{2}     
   \right] (\mathcal{H}^{\prime})^{3}
    \nonumber \\
&
 - 6 \mathcal{H}^{3}  \left[ 24K^{3} -27 \mathcal{H} K^{2} - 90  \mathcal{H}^{2} K - 12KL  + 8 \mathcal{H} L
  + 105 \mathcal{H}^{3}  - 3\mathcal{H} \left( \frac{a^{2} V_\phi}{\phi_{0}^{\prime}} \right)^{2}     
   \right] (\mathcal{H}^{\prime})^{2}
     \nonumber \\
&
 + 12  \mathcal{H}^{5}  \left[ 12K^{3} -8 \mathcal{H} K^{2} - 40  \mathcal{H}^{2} K - 6KL  + 2 \mathcal{H} L
  + 36 \mathcal{H}^{3}  - \mathcal{H} \left( \frac{a^{2} V_\phi}{\phi_{0}^{\prime}} \right)^{2}     
   \right] \mathcal{H}^{\prime}
       \nonumber \\
&
 -3  \mathcal{H}^{7}  \left[ 16 K^{3} + \mathcal{H} K^{2} - 46  \mathcal{H}^{2} K - 8KL  
  + 33 \mathcal{H}^{3}  - \mathcal{H} \left( \frac{a^{2} V_\phi}{\phi_{0}^{\prime}} \right)^{2}   \right] ,
\\
P_{7} = & \big[ 2 \mathcal{H}^{2} ( \mathcal{H}^{\prime} - \mathcal{H}^{2}) \mathcal{H}^{\prime \prime}
- 4 \mathcal{H} (\mathcal{H}^{\prime})^{3} + 2 \mathcal{H} ( \mathcal{H} K + L + 4 \mathcal{H}^{2}) 
(\mathcal{H}^{\prime})^{2}  - 4\mathcal{H}^{3} (\mathcal{H} K + L + 2 \mathcal{H}^{2}) \mathcal{H}^{\prime}
+ 2 \mathcal{H}^{5} (\mathcal{H} K + L + 2 \mathcal{H}^{2}  )   \big] 
\mathcal{H}^{\prime \prime\prime} 
 \nonumber \\
&
- 4 \mathcal{H}^{2}(\mathcal{H}^{\prime \prime})^{3} 
+ \big[ 12\mathcal{H} (\mathcal{H}^{\prime})^{2} 
- \mathcal{H} ( 8 \mathcal{H} K + 8 L - \mathcal{H}^{2} ) \mathcal{H}^{\prime}
+ \mathcal{H}^{3}(  8\mathcal{H} K + 8 L + 11 \mathcal{H}^{2}  )
  \big] (\mathcal{H}^{\prime \prime})^{2}
\nonumber \\
&
+ \big[- 9 (\mathcal{H}^{\prime})^{4} + ( 25 \mathcal{H} K + 5L - 36 \mathcal{H}^{2})
 ( \mathcal{H}^{\prime})^{3} 
 - \mathcal{H} ( 4 \mathcal{H} K^{2} + 39 \mathcal{H}^{2}K + 20 KL - 23 \mathcal{H} L
 - 54 \mathcal{H}^{3} )  (\mathcal{H}^{\prime})^{2} 
 \nonumber \\
 &
 \quad
 + \mathcal{H}^{3} (  8 \mathcal{H} K^{2} + 35 \mathcal{H}^{2} K 
 + 40 KL - 29 \mathcal{H} L 
 - 76 \mathcal{H}^{3}   ) \mathcal{H}^{\prime} 
 - \mathcal{H}^{5} (  4 \mathcal{H} K^{2} + 21 \mathcal{H}^{2}K + 20 KL
 - \mathcal{H} L - 19 \mathcal{H}^{3}  ) \big]\mathcal{H}^{\prime \prime}
\nonumber \\ 
 &
 + \left( - 19 K+ 55 \mathcal{H} + 2 \frac{a^{2} V_\phi}{\phi_{0}^{\prime}} 
    \right) (\mathcal{H}^{\prime})^{5}
 + \left[ 20 \mathcal{H} K^{2} - 6 \mathcal{H}^{2}K + 9 KL - 29 \mathcal{H} L
 -112 \mathcal{H}^{3} - 10 \mathcal{H}^{2} \frac{a^{2} V_\phi}{\phi_{0}^{\prime}} 
 - \mathcal{H} \left( \frac{a^{2} V_\phi}{\phi_{0}^{\prime}}   \right)^{2}   \right]
  (\mathcal{H}^{\prime})^{4}
\nonumber  \\
 &
 - 2\mathcal{H} \left[ 36 \mathcal{H}^{2} K^{2} - 58 \mathcal{H}^{3}K - 2L^{2} 
 + 8 K^{2}L - 27 \mathcal{H}^{2} L - 8\mathcal{H} KL-45 \mathcal{H}^{4} 
 - 10 \mathcal{H}^{3} \frac{a^{2} V_\phi}{\phi_{0}^{\prime}}
 - 2 \mathcal{H}^{2} \left( \frac{a^{2} V_\phi}{\phi_{0}^{\prime}}   \right)^{2}   \right]
  (\mathcal{H}^{\prime})^{3}
 \nonumber \\
 &
  + 2\mathcal{H}^{3} \left[ 52 \mathcal{H}^{2} K^{2} - 103 \mathcal{H}^{3}K - 6L^{2} 
 + 24 K^{2}L - 38 \mathcal{H}^{2} L - 31\mathcal{H} KL
+ 22 \mathcal{H}^{4} - 10 \mathcal{H}^{3} \frac{a^{2} V_\phi}{\phi_{0}^{\prime}}
  - 3 \mathcal{H}^{2} \left( \frac{a^{2} V_\phi}{\phi_{0}^{\prime}}   \right)^{2} 
    \right] (\mathcal{H}^{\prime})^{2}
 \nonumber \\
 &
 - \mathcal{H}^{5} \left[ 72 \mathcal{H}^{2} K^{2} - 151 \mathcal{H}^{3}K - 12L^{2} 
 + 48 K^{2}L - 74 \mathcal{H}^{2} L - 40 \mathcal{H} KL + 61 \mathcal{H}^{4} 
 - 10 \mathcal{H}^{3} \frac{a^{2} V_\phi}{\phi_{0}^{\prime}} 
 - 4 \mathcal{H}^{2} \left( \frac{a^{2} V_\phi}{\phi_{0}^{\prime}}   \right)^{2} 
     \right] \mathcal{H}^{\prime}
 \nonumber \\
 &
 + \mathcal{H}^{7} \left[ 20 \mathcal{H}^{2} K^{2} - 36 \mathcal{H}^{3}K - 4L^{2} 
 + 16 K^{2}L - 23 \mathcal{H}^{2} L - 3 \mathcal{H} KL + 16 \mathcal{H}^{4} 
 - 2 \mathcal{H}^{3} \frac{a^{2} V_\phi}{\phi_{0}^{\prime}} 
 -  \mathcal{H}^{2} \left( \frac{a^{2} V_\phi}{\phi_{0}^{\prime}}   \right)^{2}   \right] ,
\\
P_{8} = & - 6 \mathcal{H}^{2} (\mathcal{H}^{\prime} - \mathcal{H}^{2})
 \big[ \mathcal{H} \mathcal{H}^{\prime \prime}  - 4 (\mathcal{H}^{\prime})^{2}
 + 2 ( L + 3 \mathcal{H}^{2}  ) \mathcal{H}^{\prime} - 2 \mathcal{H}^{2} 
 (L + 2 \mathcal{H}^{2})   \big] \mathcal{H}^{\prime \prime\prime} 
 + 12 \mathcal{H}^{3} (\mathcal{H}^{\prime \prime})^{3}
 \nonumber \\
&
-3 \mathcal{H}^{2}  \big[ 21 (\mathcal{H}^{\prime})^{2} 
- (16 L + 17 \mathcal{H}^{2} ) \mathcal{H}^{\prime}
+ 4 \mathcal{H}^{2}(  4 L + 5 \mathcal{H}^{2}  )
  \big] (\mathcal{H}^{\prime \prime})^{2}
  \nonumber \\
  &
  + 6 \mathcal{H} \big[ 13 (\mathcal{H}^{\prime})^{4} 
- (  \mathcal{H} K + 23 L - 6 \mathcal{H}^{2}) ( \mathcal{H}^{\prime})^{3} 
 +(  8L^{2}+ 3 \mathcal{H}^{3}K + 4 \mathcal{H} KL +33 \mathcal{H}^{2} L
 - 24 \mathcal{H}^{4} )  (\mathcal{H}^{\prime})^{2} 
\nonumber \\
&
\qquad
 - \mathcal{H}^{2} (  16 L^{2}+ 3 \mathcal{H}^{3}K + 8 \mathcal{H} KL +29 \mathcal{H}^{2} L
 - 34 \mathcal{H}^{4}  ) \mathcal{H}^{\prime} 
 + \mathcal{H}^{4} ( 8 L^{2}+  \mathcal{H}^{3}K + 4 \mathcal{H} KL + 19 \mathcal{H}^{2} L
 - 5 \mathcal{H}^{4}  ) \big]\mathcal{H}^{\prime \prime} 
\nonumber \\ 
 &
 - 45 (\mathcal{H}^{\prime})^{6}
 + 6 \big[ 6 \mathcal{H} K + 11 L - 22 \mathcal{H}^{2}    \big] (\mathcal{H}^{\prime})^{5}
    \nonumber \\ 
 &
 -3 \left[7L^{2} + 2 \mathcal{H}^{2} K^{2} + 40 \mathcal{H}^{3}K + 36 \mathcal{H} KL 
 - 52 \mathcal{H}^{2} L  -152 \mathcal{H}^{4} 
 - \mathcal{H}^{2} \left( \frac{a^{2} V_\phi}{\phi_{0}^{\prime}}   \right)^{2}   \right]
  (\mathcal{H}^{\prime})^{4}
\nonumber  \\
 &
 +12\mathcal{H}  \left[ 4KL^{2} + 2 \mathcal{H}^{3}K^{2} - 4 \mathcal{H} L^{2}
- 60 \mathcal{H}^{3}L + 32 \mathcal{H}^{2} KL + 11 \mathcal{H}^{4}K - 43 \mathcal{H}^{5}
- \mathcal{H}^{3} \left( \frac{a^{2} V_\phi}{\phi_{0}^{\prime}}   \right)^{2} 
    \right] (\mathcal{H}^{\prime})^{3}
 \nonumber \\
 &
 -3 \mathcal{H}^{3}  \left[ 48 KL^{2} + 12 \mathcal{H}^{3}K^{2} - 58 \mathcal{H} L^{2}
- 356 \mathcal{H}^{3}L + 184 \mathcal{H}^{2} KL + 12 \mathcal{H}^{4}K - 41 \mathcal{H}^{5}
- 6 \mathcal{H}^{3} \left( \frac{a^{2} V_\phi}{\phi_{0}^{\prime}}   \right)^{2} 
    \right] (\mathcal{H}^{\prime})^{2}
 \nonumber \\
 &
 + 6 \mathcal{H}^{5}  \left[ 24 KL^{2} + 4 \mathcal{H}^{3}K^{2} - 20 \mathcal{H} L^{2}
- 123 \mathcal{H}^{3}L + 64 \mathcal{H}^{2} KL - 4 \mathcal{H}^{4}K + 6 \mathcal{H}^{5}
- 2 \mathcal{H}^{3} \left( \frac{a^{2} V_\phi}{\phi_{0}^{\prime}}   \right)^{2} 
    \right] \mathcal{H}^{\prime}
\nonumber \\
 &
 -3 \mathcal{H}^{7}  \left[ 16 KL^{2} + 2 \mathcal{H}^{3}K^{2} - 5 \mathcal{H} L^{2}
- 56 \mathcal{H}^{3}L + 36 \mathcal{H}^{2} KL - 4 \mathcal{H}^{4}K + 6 \mathcal{H}^{5}
-  \mathcal{H}^{3} \left( \frac{a^{2} V_\phi}{\phi_{0}^{\prime}}   \right)^{2} 
    \right] .
\end{align}

\end{document}